\documentclass[conference]{IEEEtran}
\IEEEoverridecommandlockouts
\usepackage{times}
\usepackage{cite}
\usepackage{amsmath,amssymb,amsfonts}
\usepackage{graphicx}
\usepackage{booktabs}
\usepackage{xcolor}
\usepackage{algorithm}
\usepackage{algpseudocode}

\begin{document}

\title{Silent Blindness: Fault Vulnerability of Hardware Malware Detectors}

\author{
\IEEEauthorblockN{\small Alireza Abolhasani Zeraatkar$^{1}$, Parnian Shabani Kamran$^{1}$, Parsa Mirfasihi$^{2}$, Saleh Khalaj Monfared$^{3}$, Hussain Al-Asaad$^{1}$}
\IEEEauthorblockA{
$^{1}$Electrical and Computer Engineering Department, University of California, Davis, United States\\
$^{2}$Electrical and Computer Engineering Department, San Francisco State University, San Francisco, United States\\
$^{3}$Electrical and Computer Engineering Department, Worcester Polytechnic Institute, Worcester, United States\\
Email: \{abolhasani, pkamran\}@ucdavis.edu, smirfasihi@sfsu.edu, skmonfared@wpi.edu, hsalasaad@ucdavis.edu}
}

\maketitle
\thispagestyle{empty}
\pagestyle{empty}

\begin{abstract}
Hardware malware detectors (HMDs) classify hardware performance-counter measurements to identify malicious execution. Faults in their decision logic or parameters can weaken detection and allow malware to be classified as benign. We investigate this vulnerability through structural fault injection across a design space comprising 18 classifiers, four counter budgets, and AdaBoost and Bagging ensembles. The simulation evaluates 453,174 decision stuck-at and selected parameter bit-flip events on a 7,400-instance. Primary comparisons use 128 completed configurations. These completed-subset results are distinguished from the simulation summaries covering 140 configurations with enumerable fault sites. Fault outcomes are classified by their security consequences, including evasion, false-alarm increases, and metric silence. A fault is critical when malware recall decreases by at least five percentage points. Tree-based models exhibit a lower mean critical-fault fraction than function-based models, 5.3\% versus 14.3\%; exhaustive-only analysis preserves this ordering. Moreover, 72.3\% of evaluated tree faults are metric-silent, changing both recall and false-positive rate by less than 0.1 percentage points without necessarily preserving individual predictions. Among the 27 completed matched comparisons, Bagging reduces criticality in 26, whereas AdaBoost increases it in two. Finally, an exact minimax selection method identifies protection candidates under additive unit costs and ideal correction. For a representative J48 detector, protecting 16 of 115 structural units reduces predicted maximum recall loss from 60.1 to 3.3 percentage points. These results characterize structural-model sensitivity and potential protection benefits; faulted RTL correspondence and physical hardening overhead remain unvalidated.
\end{abstract}


\section{Introduction}

A hardware malware detector (HMD) is a classifier that operates at the hardware
level: it monitors the hardware performance counters (HPCs) of running programs
and flags suspicious activity. It is a lightweight component of security
silicon, and the system trusts it to keep monitoring. That trust makes the
detector itself a target and turns its failure modes into a security concern
rather than a minor reliability issue. A degraded branch predictor merely slows
a processor, whereas a faulty malware detector can let malware execute while
classifying its behavior as benign.

This study examines how failures in individual decision points and stored
parameters, whether caused by manufacturing defects, aging, or deliberate fault
injection, affect deployed HMDs. Rather than focusing on a single prototype, the
analysis covers an entire published detector design space. The dataset is the
144 trained detector configurations of~\cite{zeraatkar2024}, 18 classifiers over
16-, 8-, 4-, and 2-counter feature sets, plus AdaBoost and Bagging ensembles.
Their hardware implementations were validated bit-exactly against the
corresponding trained models in~\cite{paper1}. We inject two hardware-relevant
fault models into the 140 configurations with enumerable fault sites.

Our analysis is guided by the principle that a fault in a \emph{detector} has
security implications beyond functional correctness. Mean accuracy degradation,
the conventional evaluation metric, obscures three distinct outcomes that
require different defensive responses. An \emph{evasion} fault reduces recall
and lets malware execute undetected, directly serving the attacker's objective.
A \emph{denial-of-service} fault raises the false-positive rate enough that
operators stop trusting the detector and may disable it. A \emph{silent} fault
leaves both recall and false-positive rate unchanged, so output monitoring alone
cannot distinguish the defective detector from a healthy one.

Our measurements show that these outcomes are not evenly distributed and that
their distribution depends on the classifier's dominant hardware primitive:
comparators or multiply-accumulate units. In a 4-HPC J48 decision tree, only
$3.9\%$ of the $3{,}910$ fault sites cause evasion, while $51.5\%$ are silent;
the worst single fault drops malware recall from $97.0\%$ to $36.9\%$. In a
4-HPC SGD detector, the picture inverts. The design exposes only $25$ fault
sites, but $28\%$ of them cause evasion and a further $44\%$ cause denial of
service. None are silent, and a single bit flip drives recall to $0\%$.
Comparator-based detectors spread their decision logic across many nodes,
whereas multiply-accumulate detectors concentrate it in a few weights, any one
of which, if faulted, can cause severe failure.

This study addresses five research questions:

\textbf{RQ1:} For each classifier family (trees, rules, function-based models,
and Bayesian models), what fraction of faults is security-critical? Is that
fraction governed more by the classifier's dominant hardware primitive,
comparators or multiply-accumulate units, than by its number of fault sites?

\textbf{RQ2:} How much damage can a detector sustain while its recall and
false-positive rate remain indistinguishable from those of a healthy detector?
Where in the detector's structure do the damaging faults lie?

\textbf{RQ3:} When the feature set is reduced from 16 to 8, 4, and 2 counters
to lower silicon cost, how does each classifier's fraction of critical faults
change?

\textbf{RQ4:} At the same HPC budget, do AdaBoost and Bagging ensembles, which
add little detection accuracy per unit area, reduce the fraction of critical
faults? Does equal-weight Bagging tolerate faults better than weighted boosting?

\textbf{RQ5:} Can selective hardening guided by fault criticality bound the
worst-case single-fault recall loss at a lower area cost than full redundancy?
How large a hardening budget does that bound require?

This study makes five main contributions:
\begin{itemize}
\item It introduces a fault taxonomy based on security consequences. Each
injected fault is classified as evasion, denial of service, silent, noisy, or
degraded according to its effect on the defender rather than by its mean
accuracy loss.
\item It provides a comprehensive fault analysis of a published HMD design
space. Decision stuck-at faults and parameter bit flips are injected into 140
validated configurations covering the tree, rule, function, and Bayesian
classifier families, and the analysis quantifies how counter-budget reduction
and ensembling change fault vulnerability.
\item It identifies a size-density inversion in the attack surface.
Comparator-based detectors expose far more fault sites than
multiply-accumulate detectors, yet a smaller fraction of their faults is
critical, so fault-site count is a poor proxy for vulnerability. The mean critical-fault fraction differs across classifier families, among completed configurations, it is approximately 2.5 times higher for function-based detectors than for tree-based detectors (14.3\% versus 5.7\%).

\item It quantifies silent faults, which leave both recall and false-positive
rate unchanged and therefore cannot be detected by output monitoring alone, and
it identifies the structural regions in which critical faults concentrate.
\item It proposes selective hardening guided by the concentration of critical
faults in a small subset of structural units. It provides an exact minimax
selection algorithm under the assumptions of disjoint units, additive costs, and
ideal correction, and it evaluates the predicted trade-off between normalized
protection cost and residual worst-case recall loss over the evaluated fault
set.
\end{itemize}

\section{Related Work}
\label{sec:related}

\textbf{Hardware-assisted malware detection.} Detecting malware from performance counters was shown feasible by Demme et al.~\cite{demme2013} and researchers focused on making this approach practical by integrating the classifier into the pipeline~\cite{ozsoy2015,ozsoy2016}, replacing single learners with ensembles~\cite{khasawneh2015,khasawneh2020ehmd}, introducing programmable monitoring hardware~\cite{delshadtehrani2020}, customising the counter set to the workload~\cite{kuruvila2021}, reducing the classifier to the two to four counters that a practical performance-monitoring unit can sample concurrently~\cite{kadiyala2020,makrani2022optiedge,gao2021adaptive}, improving the feature set against noisy or evasive traces~\cite{sayadi2021stealth,he2024deephmd,he2026transformers,agredo2025b}, adapting detection cost to workload with reinforcement learning~\cite{he2022reinforced}, and coupling the numeric decision to a semantic explanation~\cite{golzari2026}. Surveys collect the resulting design space~\cite{chenet2024,sayadi2024}, comparative studies rank classifier families by detection quality~\cite{zeraatkar2024}, field deployments identify which methods remain practical~\cite{konstantinou2022}, and a companion study assesses those rankings in routed silicon~\cite{paper1}. However, none of this literature examines how a detector behaves once its own logic is damaged, which is the main question addressed in this study.

\textbf{Ransomware-specific hardware detection.} A parallel line of work applies HPC-based detection specifically to ransomware, where the file encryption leaves a distinctive microarchitectural signature that could be detected before damage. This approach includes early-warning classifiers that flag encryption activity within seconds of payload execution~\cite{anand2023}, imaging-style CNN detectors that generalize across unseen families~\cite{ganfure2023}, lightweight monitors for IoT and edge devices~\cite{mofidi2025,radhakrishna2024edge}, classifiers tested against obfuscated and packed variants~\cite{majd2023ransomware,cassel2023ransomware,cassel2024iot}, and mechanisms that intervene directly in the encryption pipeline rather than passive flagging ~\cite{alam2020}. These detectors use the same implementation logic as general-purpose HMDs and are therefore exposed to the same fault surface this study presents.

\textbf{Attacking the detector.} The adversarial line of work treats an HMD as a classifier to be misled. Zhou et al.~\cite{zhou2018,zhou2021} and Kazdagli et al.~\cite{kazdagli2016} question how much reported accuracy survives realistic evaluation. Basu et al.~\cite{basu2026} bound what sampled counters can detect at all. Islam et al.~\cite{islam2022} reverse-engineer a deployed detector and evade it cheaply, motivating the randomized switching of RHMD~\cite{khasawneh2017rhmd} and its ensemble counterpart EnsembleHMD~\cite{khasawneh2020ehmd}. Obfuscation-aware~\cite{he2024obfusgate} and reinforcement-learning-hardened~\cite{he2024drl} detectors push back against the same evasion threat from the defender's side. All these attacks and their corresponding defenses operate through the detector's \emph{inputs} and assume its logic is intact and correct, an assumption that applies to cost-adaptive designs that power the classifier down and back up on suspicion~\cite{zeraatkar2026wake}. In this paper, we are interested in evaluating the vulnerability of the implementation in the presence of faults, asking what an adversary gains by compromising the HMD's logic. 

\textbf{Fault injection as a realistic option.} Voltage manipulation can be localized to direct faults to the selected regions of a die~\cite{khojasteh2026}. Undervolting can force a chip into a state-retaining sleep mode that disables its internal sensors~\cite{mitard2025}. Impedance analysis extracts data bits without invasive access~\cite{monfared2023} and can also be used to detect tampering~\cite{mosavirik2023}. A hypervisor-level debugger makes the software component of such simulations relatively straightforward~\cite{karvandi2022}.

\textbf{Positioning against fault injection on machine-learning hardware.} Fault injection against machine-learning hardware is a well-studied line of work for deep neural networks. The attacker flips a few of weight bits in a quantized DNN's DRAM parameters to break its accuracy~\cite{rakin2019bfa}. Once triggered through Rowhammer, these changes can be introduced deterministically ~\cite{yao2020deephammer}. The resulting defenses include integrity hashing~\cite{javaheripi2021hashtag}, criticality-guided error correction codes (ECCs) ~\cite{traiola2023hardnning}, and selective triple-modular redundancy of only the most vulnerable layers~\cite{libano2019selective}. All focus on the same target, a large weight memory in which changing only a few bits is detrimental. A recent systematic review confirms this holds across DNN accelerators~\cite{ahmadilivani2023survey}. Hardware malware detectors invert that picture. Most of the families in our design space, decision trees, rule lists, and their ensembles, have no weight memory to attack at all. Their decision logic is combinational comparator and gating logic realized directly in the netlist. Consequently, the DNN literature's attack-and-defense strategy, which locates the vulnerable memory bits and then protects them, cannot be applied directly. However, its underlying principle remains relevant. We adapt the criticality-guided selective protection in Section~\ref{sec:harden} to prevent the attacker from compromising the HMD. To our knowledge, this is the first fault analysis of HPC-based malware detectors, the first at the scale of a complete HMD design space.

 \section{Threat and Fault Model}
\label{sec:threat}

\subsection{Fault models}

We inject two types of faults, each representing a physical failure mechanism that could plausibly result from either a deliberate attack or a defect in a deployed HMD.

\textbf{Decision stuck-at.} A comparator output, rule-condition result, table
entry, or ensemble member verdict is forced permanently to one value. This is
the classical stuck-at-0, stuck-at-1 model of manufacturing test, and it also captures the
observable behavior of a shorted or open net, a latched glitch, and an aged
device that no longer switches. In a comparator-based detector the decision
points \emph{are} the computation, so this model localizes exactly what an
adversary would want to control.

\textbf{Parameter bit-flip.} A single bit of a stored parameter is inverted, whether a tree threshold, a rule bound, a table interval, a linear or MLP weight, or an ensemble vote weight. Integer thresholds are flipped across their 32-bit
representation, matching the unsigned datapath of the implementations under
study; real-valued weights are flipped in the mantissa, exponent, and sign
positions of their IEEE-754 encoding. This model captures transient upsets,
faulty parameter memory, and targeted injection against a detector whose
coefficients are stored rather than hard-wired.

\subsection{Adversary}

We assume an adversary whose objective is to execute malware without detection and
who can induce a limited number of faults in the detector but cannot completely
rewrite its logic. This capability is well established, as supply-voltage
and clock manipulation, electromagnetic and laser injection, and undervolting
attacks that disable on-chip sensors have all been demonstrated against fabricated
devices \cite{khojasteh2026,mitard2025}. We deliberately do
\emph{not} assume the ability to read the detector's parameters, and we treat
the number of faults an adversary must place as the cost metric.

A second, adversary-free reading of the same experiments is equally important. A
manufacturing defect or an aging-induced failure produces the same fault states
with no attacker at all. Under that reading, the silent-fault fraction indicates
how much of a detector can break without any output-level test noticing.

\section{Methodology}
\label{sec:method}

\subsection{Detectors and validated baselines}

The experiment is the 144 trained detector configurations of \cite{zeraatkar2024},
in the validated hardware form given to them by \cite{paper1}. It spans 18 classifiers from the tree, rule, function-based (linear and MLP), Bayesian, and instance-based families, each trained under 16-, 8-, 4-, and 2-counter budgets, plus
AdaBoost and Bagging ensembles of $K{=}10$ members at the two reduced budgets.
Crucially, each configuration includes a software model
that was validated against the recorded per-instance predictions of the original
trained classifiers and a synthesizable RTL that was simulated against that
model over the full dataset with zero mismatches. Faults are therefore injected into a representation whose correct behaviour has already been verified against the deployed detector. Any measured change in behaviour can thus be attributed solely to the injected fault.

\subsection{Injection level}

Faults are injected into the structural model of each detector, its comparator tree, rule list, table, weight vector, or ensemble, rather than into a gate netlist. This choice is deliberate and has two motivations. First,
\emph{interpretability}, a fault at this level names the architectural element
that failed (``the root comparator of member 7''), which is what a hardening
decision needs, whereas a gate-level net identifier is not portable across
synthesis runs or technology targets. Second, \emph{portability}, the same fault
site has a meaning on both technology targets of \cite{paper1} and in any future
implementation of the same classifier, so vulnerability profiles are properties
of the detector architecture rather than of one synthesis result. The
corresponding limitation is that our analysis considers architectural fault sites  rather than
physical ones. A gate-level simulation would also cover faults introduced by
synthesis. Establishing agreement between the architectural and physical levels is left for future work.

\subsection{Evaluation convention}

Each fault-injected detector is evaluated on the same $7{,}400$-instance used for baseline validation. Changes in recall and false-positive rate are measured relative to that detector’s fault-free baseline on the same inputs. We use the full study to maintain consistency with the archived per-instance prediction records. This paired evaluation characterizes fault-induced changes in detector behavior on the available study; it does not estimate detection performance on unseen programs. Held-out detection performance is reported separately in the companion study \cite{paper1}.

\subsection{Security-semantic classification}
For each fault, we recompute recall $R$, the fraction of malware detected, and
false-positive rate $F$, the fraction of benign applications flagged, and
compare them with the fault-free baseline $(R_0, F_0)$ of the same detector.
The classification uses three parameters, an evasion threshold $\tau_R$ on the
loss of recall, a denial-of-service threshold $\tau_F$ on the rise in
false-positive rate, and a tolerance $\epsilon$ below which a change is treated
as zero. Each fault is assigned the first of the following classes whose
condition it meets. It is \textbf{noisy} if the faulted detector fails to
produce a well-formed decision, which in silicon corresponds to a malformed
output that a wrapper-level check can detect; \textbf{evasion} if
$R_0 - R \ge \tau_R$, meaning that the detector now misses malware it
previously caught; \textbf{denial-of-service} if $F - F_0 \ge \tau_F$, meaning
that it now floods the verification stage with false alarms; \textbf{silent} if
$|R - R_0| < \epsilon$ and $|F - F_0| < \epsilon$, meaning that the fault is
present but no output-level metric reveals it; and \textbf{degraded} otherwise.
Evasion precedes denial of service because a fault that causes both still lets
malware through.

This taxonomy specializes the failure-mode classification of Avi\v{z}ienis et
al.~\cite{halting2004}, which grades failures by detectability and consequence.
Noisy faults are \emph{signaled} failures; evasion and denial-of-service faults
are \emph{unsignaled} content failures of differing consequence; and silent
faults are those whose errors remain latent or too small to observe at the
detector's output. Our \emph{silent} therefore differs from the \emph{silent
failure} of~\cite{halting2004}, in which a component stops delivering output
altogether.

We set $\tau_R = \tau_F = 5$ percentage points (pp) and $\epsilon = 0.1$~pp
throughout. The tolerance $\epsilon$ sits just above the resolution, where $0.1$~pp corresponds to fewer than $n_\epsilon$ instances
of the smaller class, and far below what an output monitor could observe:
detecting a shift of $\epsilon$ in a rate $p$ with a one-sample test at
$\alpha = 0.05$ and $80\%$ power requires $n \approx 7.8\,p(1-p)/\epsilon^2$
labeled observations, more than $7 \times 10^4$ for any baseline rate between
$1\%$ and $99\%$, and labels are precisely what an in-field monitor lacks. The
thresholds $\tau_R$ and $\tau_F$ are set at an operationally significant level:
a $5$~pp loss of recall lets one additional malware sample in twenty execute
undetected, comparable to the entire gain in detection performance that a
$K{=}10$ ensemble provides over its base learner (one to six
points~\cite{paper1}). Using one value for both thresholds places evasion and
denial of service on a common scale, so that the asymmetry between them lies
solely in which class we treat as critical.

We call a fault \emph{critical} when it is an evasion fault, the class that
defeats detection itself, and report denial-of-service faults separately. The
asymmetry is intentional, in the deployment architecture we assume, the HMD
escalates flagged programs to a software verification stage, so a missed
detection is a security failure whereas a false alarm is a cost. Because
criticality depends only on $\tau_R$, we test its sensitivity by sweeping
$\tau_R$ from $1$ to $10$~pp. The family ordering is unchanged across the
sweep, the $95\%$ confidence intervals of the tree and function families'
critical-fault fractions remain disjoint at every threshold.

\section{Results}
\label{sec:results}

\begin{table}[t]
\caption{Enumerated fault-site totals and mean per-configuration outcome percentages across the 128 completed configurations. The 12 time-limited configurations are excluded. Intervals are 95\% normal confidence intervals across configurations. The KNN row covers ensemble vote weights only.}
\label{tab:family}
\centering
\footnotesize
\setlength{\tabcolsep}{2.5pt}
\begin{tabular}{@{}lrrrrr@{}}
\toprule
Family & Designs & Fault sites & Evasion (\%) & DoS (\%) & Silent (\%) \\
\midrule
Trees & 52 & 866,056 & $5.7\pm2.7$ & 2.5 & $70.2\pm5.6$ \\
Rules & 26 & 375,109 & $5.4\pm2.5$ & 3.2 & $65.4\pm7.2$ \\
Function-based & 40 & 12,410 & $14.3\pm4.3$ & 16.3 & $30.9\pm8.8$ \\
Bayes & 6 & 18,100 & $4.5\pm2.5$ & 3.7 & $53.2\pm17.7$ \\
KNN votes & 4 & 110 & $10.0\pm11.3$ & 0.0 & $90.0\pm11.3$ \\
\bottomrule
\end{tabular}
\end{table}

\begin{table}[t]
\caption{Observed critical-fault percentages at four HPCs for classifier triples in which the single detector and both ensembles completed their evaluations. Percentages are rounded to two decimal places. Zero denotes no critical faults observed in the evaluated set, rather than a guarantee over all possible faults.}
\label{tab:ensemble}
\centering
\footnotesize
\setlength{\tabcolsep}{4pt}
\begin{tabular}{@{}lrrr@{}}
\toprule
Classifier (4 HPCs) & Single (\%) & AdaBoost (\%) & Bagging (\%) \\
\midrule
SMO & 36.00 & 8.00 & 0.00 \\
Decision Stump & 35.29 & 21.79 & 0.00 \\
SGD & 28.00 & 4.67 & 0.00 \\
Logistic & 24.00 & 8.89 & 0.00 \\
MLP & 22.58 & 0.09 & 0.00 \\
Simple Logistic & 10.00 & 1.20 & 0.00 \\
REPTree & 6.77 & 0.05 & 0.00 \\
PART & 6.37 & 0.00 & 0.00 \\
JRip & 5.70 & 1.18 & 0.00 \\
J48 & 3.94 & 0.01 & 0.00 \\
Hoeffding Tree & 2.94 & 4.11 & 0.00 \\
Random Tree & 1.93 & 1.85 & 0.00 \\
LMT & 1.85 & 0.00 & 0.00 \\
\bottomrule
\end{tabular}
\end{table}

\begin{table}[t]
\caption{Study size and coverage within the enumerated fault model. Exhaustive configurations have zero sampling uncertainty for their enumerated fault sets. The nominal sampling margin applies only to completed random samples; no campaign-wide sampling margin is asserted.}
\label{tab:coverage}
\centering
\footnotesize
\setlength{\tabcolsep}{4pt}
\begin{tabular}{@{}lr@{}}
\toprule
Quantity & Value \\
\midrule
Configurations in the summary & 144 \\
Zero-site single KNN configurations & 4 \\
Configurations with enumerable fault sites & 140 \\
Primary completed configurations & 128 \\
Fault sites enumerated across the full campaign & 5,585,493 \\
Fault evaluations across the full campaign & 453,174 \\
Corpus evaluations across the full campaign & $3.35\times10^{9}$ \\
Exhaustively evaluated configurations & 101 \\
Completed randomly sampled configurations & 27 \\
Time-limited configurations & 12 \\
Maximum nominal 95\% margin, completed random samples & 1.0\,pp \\
study-wide sampling margin & Not asserted \\
\bottomrule
\end{tabular}
\end{table}

\subsection{Attack surface: size versus density (RQ1)}

\begin{figure}[t]
\centering
\includegraphics[width=\columnwidth]{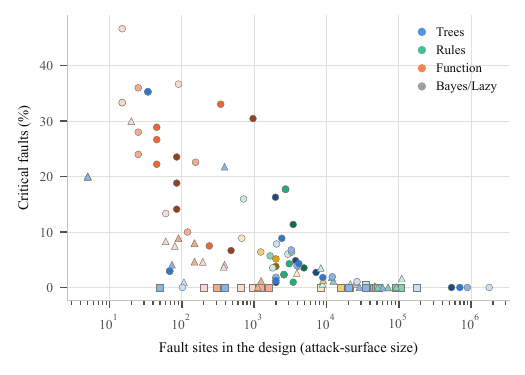}

\caption{Enumerated fault-site count versus critical-fault percentage for each detector configuration.}

\label{fig:inversion}
\end{figure}

Table~\ref{tab:family} and Fig.~\ref{fig:inversion} summarize the simulation across 140 configurations with enumerable fault sites, whereas the primary analysis reported in the abstract uses the 128 completed configurations. The two axes of vulnerability move in opposite directions.
Tree-based detectors expose $4.77$ million fault sites against the function
family's $12{,}410$, a factor of $384$, yet only $5.3\%$ of tree faults are
critical against $14.3\%$ for function detectors. Comparator-based detectors
distribute their decision logic over many individually inert nodes, whereas
multiply-accumulate detectors concentrate it into a handful of weights each of
which is load-bearing. An adversary attacking a tree has vastly more places to
strike and a much lower chance that any given strike matters; an adversary
attacking a linear detector has few targets and can hardly miss.

This inverts the ranking of the companion cost study, the detectors that are
cheapest in silicon are \emph{not} the most fragile per fault site, but they do
present the largest surface, and, as the next subsection shows, they hide the
most damage.

\subsection{The silent majority (RQ2)}

\begin{figure}[t]
\centering
\includegraphics[width=\columnwidth]{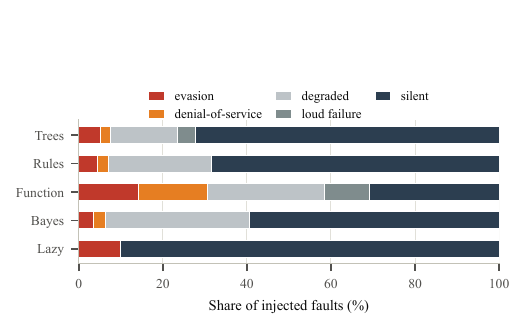}
\caption{Fault-outcome distributions by classifier family.}
\label{fig:verdicts}
\end{figure}

Fig.~\ref{fig:verdicts} decomposes every family's faults by security outcome.
In the simulation summary, the mean metric-silent fraction is 72.3\% for tree detectors and 68.4\% for rule detectors, using a tolerance of 0.1 percentage points for changes in both recall and false-positive rate. Such a detector is defective and no output-level check, neither a watchdog on the decision nor a plausibility bound on the alarm rate, can distinguish it from a healthy one. The function family is
the exception that proves the rule, with only a few load-bearing weights, its
faults are three times less likely to be silent and far more likely to be noisy,
because a faulty weight typically produces an out-of-range accumulation that
the datapath itself rejects.

Critical and metric-silent faults are disjoint under the stated thresholds: critical faults reduce aggregate recall by at least five percentage points, whereas metric-silent faults change recall and false-positive rate by less than 0.1 percentage points. Metric silence can nevertheless conceal changes in individual predictions or malware-subclass recall. Selective hardening therefore considers fault detection alongside correction.

\subsection{Where criticality lives (RQ2)}

\begin{figure}[t]
\centering
\includegraphics[width=\columnwidth]{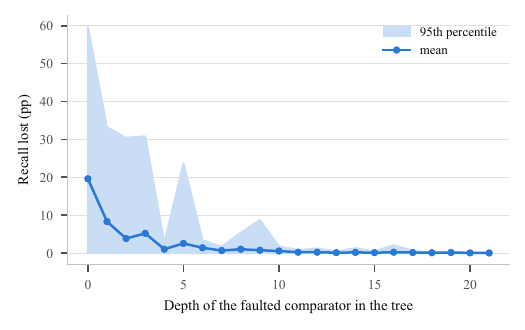}
\caption{Mean and 95th-percentile recall loss by comparator depth across J48 and REPTree configurations.}
\label{fig:depth}
\end{figure}

Fig.~\ref{fig:depth} resolves the damage of tree faults by the depth of the
faulted comparator. Critical faults are strongly concentrated near the root. On average, a root fault reduces recall by $19.5$ percentage points, with a maximum reduction of $60$ points. In contrast, faults below depth ten have almost no effect on mean recall. The distribution is heavy-tailed rather than uniform, which makes selective
hardening viable because protecting a small number of shallow comparators removes most of
the exploitable surface, and Section~\ref{sec:harden} quantifies exactly how much.

\subsection{The counter budget trades silicon for fragility (RQ3)}

Reducing the counter budget is the cheapest way to shrink an HMD, and the
companion cost study recommends it, four counters retain essentially all of the
sixteen-counter detection performance at roughly half the area. The fault
simulation shows that this saving is not free in the fault domain. Among function-based detectors, reducing the counter budget from sixteen to two increases the critical-fault fraction from 18.8\% to 46.7\% for SGD, from 23.5\% to 33.3\% for SMO, and from 14.1\% to 33.3\% for Logistic. However, the trend is not necessarily monotonic: SMO criticality decreases from 36.0\% at four counters to 33.3\% at two. With fewer counters, each surviving weight can carry a larger share of the decision, increasing the potential impact of a fault.

The practical reading is that counter reduction and fault robustness pull in
opposite directions, and that the pull is far stronger for detectors whose
decision is concentrated in a few coefficients. A designer who adopts the
four-counter operating point on cost grounds should expect a modest increase in
per-fault criticality for tree and rule detectors, and a substantial one for
linear detectors, which is a further argument, on top of the cost argument, for
preferring the comparator-based families at reduced budgets.

\subsection{Ensembles buy redundancy, not accuracy (RQ4)}

\begin{figure*}[t]
\centering
\includegraphics[width=\textwidth]{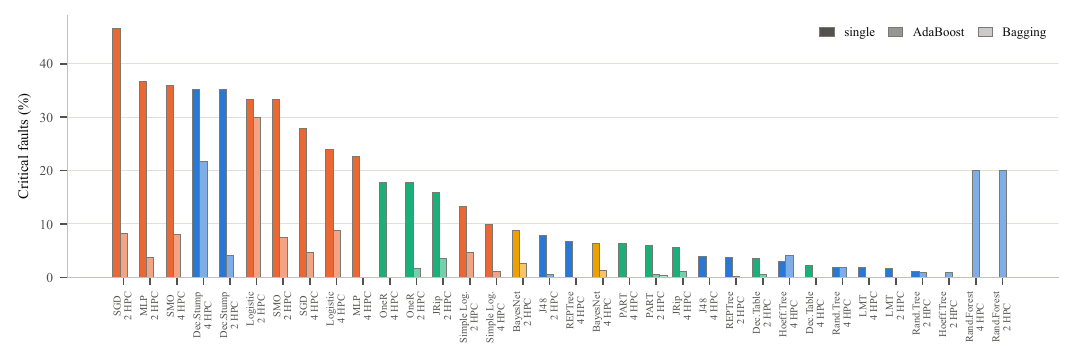}
\caption{Critical-fault percentages for single detectors, AdaBoost, and Bagging at two and four HPCs.}

\label{fig:ensemble}
\end{figure*}

The companion cost study concludes that ensembling is poor value, a $K{=}10$
ensemble costs ten times the silicon of its base learner and buys one to six
points of detection performance. Fig.~\ref{fig:ensemble} and
Table~\ref{tab:ensemble} show that the same silicon buys something the accuracy
analysis cannot see. Across the 34 classifier/budget triples, Bagging reduced the critical-fault fraction in 31 and left it unchanged in three. Bagging achieved zero criticality in 33 comparisons, including all 17 at four HPCs. AdaBoost increased criticality in four comparisons.

Equal-weight Bagging is the stronger of the two, and the reason is structural:
with ten equally weighted votes, no single faulty member can move the
majority, so the critical-fault fraction falls to $0.0\%$ for almost every base
learner. AdaBoost is weaker because its members are weighted unequally, the
vote of a high-weight member can dominate, so fault injection in it still moves the
decision, and the residual critical-fault fraction ranges from 0\% to 30\%.
The most striking case is the multilayer perceptron, as a single detector
$22.6\%$ of its faults are critical and the worst drives recall to $0\%$; as a
bagged ensemble no single fault is critical and the worst-case recall never
falls below $63.6\%$.

This is the practical reconciliation of the two studies. Where silicon is the
binding constraint and the threat model is benign, the single-tree detector of
the cost study remains the right choice. Where the detector must keep working under faults, in a high-assurance deployment, a long-lived part, or against an adversary with physical access, the ensemble's redundancy is not wasted silicon but the
mechanism that provides fault tolerance, and it should be selected for that
reason rather than for its detection score.

\section{Criticality-Guided Selective Hardening (RQ5)}
\label{sec:harden}

The simulation of Section~\ref{sec:method} shows that critical faults are rare and
structurally concentrated. That invites a defense that is far cheaper than
traditional redundancy which protect only what matters. Turning that intuition into an
algorithm requires three observations.

\ A hardening mechanism cannot
protect ``bit 23 of the threshold at node 5''; it protects a \emph{structural unit}, a comparator together with the threshold register that feeds it, a
weight register, an ensemble member. Each unit therefore covers a set of fault
sites, which makes selection a coverage problem rather than a ranking problem.

Detection is cheaper than correction, and often sufficient. Triple
modular redundancy masks a fault at roughly $2.3\times$ the protected unit's
area; duplicate-and-compare (or parity on a stored parameter) only
\emph{detects} it, at roughly $1.15\times$. For a security monitor, however,
detection is not a weak substitute, a detector that knows it is damaged can
raise an integrity flag and fall back to software scanning, which is a safe
state. The failure mode that must be eliminated is not damage but \emph{silence}.

The adversarial and stochastic objectives differ. An attacker places
the single worst fault they can find, so the security objective is to minimize
the \emph{worst} residual damage. A defect or an upset strikes at random, so the
reliability objective is to minimize the \emph{expected} damage. These select
different units, and we report both.

Algorithm~\ref{alg:cgsh} states the procedure. The adversarial objective admits
an \emph{exact} solution, for a candidate residual bound $t$, the set of units that must be hardened is forced, namely every unit whose worst fault exceeds $t$, and that set's cost is monotone in $t$, so binary search over the observed damage
levels finds the minimum achievable worst-case residual under the budget with no
approximation. The stochastic objective is budgeted maximum coverage, which is
NP-hard; the greedy marginal-gain-per-cost rule with the single-best-item
comparison retains the standard $(1-1/e)$ guarantee.

\begin{algorithm}[t]
\caption{Criticality-guided selection of structural units for fault correction and detection.}

\label{alg:cgsh}
\begin{algorithmic}[1]
\Require fault simulation $F$; structural units $U$; unit areas $a(\cdot)$;
         budget $B$; mechanism costs $\kappa_{\mathrm{det}}, \kappa_{\mathrm{cor}}$
\Ensure hardening assignment $A: U \to \{\text{none}, \text{detect}, \text{correct}\}$
\State \textbf{aggregate} for each $u \in U$:
       $w_{\max}(u) \gets \max_{f \in u} \Delta R(f)$,
       $w_{\Sigma}(u) \gets \sum_{f \in u} \Delta R(f)$,
       $s(u) \gets |\{f \in u: f \text{ silent}\}|$
\Statex \textbf{(i) adversarial tier: exact minimax}
\State $L \gets \mathrm{sort}(\{0\} \cup \{w_{\max}(u)\})$;\;
       $lo \gets 0$;\; $hi \gets |L|-1$;\; $C \gets \emptyset$
\While{$lo \le hi$}
  \State $t \gets L[\lfloor (lo+hi)/2 \rfloor]$;\;
         $N \gets \{u: w_{\max}(u) > t\}$
  \If{$\kappa_{\mathrm{cor}} \sum_{u \in N} a(u) \le \beta B$}
     \State $C \gets N$;\; $t^{*} \gets t$;\; $hi \gets \lfloor (lo+hi)/2 \rfloor - 1$
  \Else
     \State $lo \gets \lfloor (lo+hi)/2 \rfloor + 1$
  \EndIf
\EndWhile
\Statex \textbf{(ii) detection tier: eliminate silence with what is left}
\State $B' \gets B - \kappa_{\mathrm{cor}}\sum_{u \in C} a(u)$;\; $D \gets \emptyset$
\ForAll{$u \in U \setminus C$ in decreasing $s(u) / (\kappa_{\mathrm{det}} a(u))$}
  \If{$s(u) > 0$ \textbf{and} $\kappa_{\mathrm{det}} a(u) \le B'$}
     \State $D \gets D \cup \{u\}$;\;
            $B' \gets B' - \kappa_{\mathrm{det}} a(u)$
  \EndIf
\EndFor
\Statex \textbf{(iii) verify empirically, do not trust the prediction}
\State emit hardened RTL for $A$; \textbf{re-run} the fault simulation on it
\State \Return $A$, worst-case residual, silent fraction
\end{algorithmic}
\end{algorithm}

Two properties of the procedure matter for the evaluation. First, the selection
is driven by \emph{measured} damage, so we can compare it against the structural
heuristic that a designer would reach for without a fault simulation, namely protecting the top $d$ levels of the tree, and report whether the simulation was necessary or whether depth is a sufficient proxy. Second, step~(iii) re-runs the simulation
on the hardened design rather than reporting the residual risk that selection
predicted; predicted and achieved are reported side by side, and the hardened
variants are carried through the same open-PDK flow as the originals so that the
budget axis is measured silicon rather than an estimate. The sweep reported
below uses the area apportionment of Section~\ref{sec:harden}; carrying the
hardened variants through the physical-design flow, and thus replacing the
apportioned budget axis with measured silicon, is the natural next step.

\begin{figure}[t]
\centering
\includegraphics[width=\columnwidth]{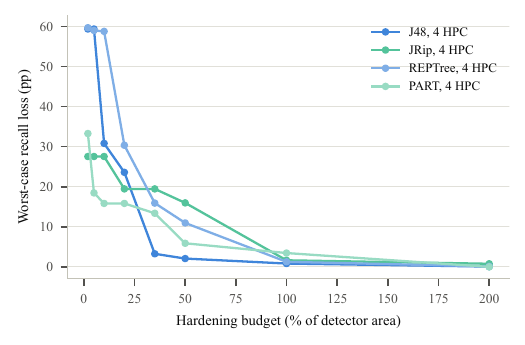}
\caption{Predicted maximum recall loss over evaluated faults versus estimated hardening budget, expressed as a percentage of baseline detector area.}
\label{fig:harden}
\end{figure}

Fig.~\ref{fig:harden} sweeps the budget for four representative detectors. The
curves are steep and then flat, which is the practical message, for the
recommended J48 detector, a hardening budget of $35\%$ of the detector's own
area reduces the worst single-fault damage from $60.1$ percentage points of
recall to $3.3$, and beyond that point the returns are negligible. Protecting
16 of the detector's 115 structural units captures almost the entire
exploitable surface.

\section{Limitations}
\label{sec:limits}

Three limitations bound the scope of these results. \emph{Injection level}, faults
are injected into the structural model of each detector rather than into a gate
netlist, which makes the fault sites architectural and portable but excludes
faults introduced by synthesis itself; we expect the two levels to agree on the
family ordering, but we have not shown it. \emph{Coverage of instance-based
detectors}, the KNN configurations store their training set rather than
parameters, and our site enumeration does not yet address stored-sample
fault, so the KNN rows of Table~\ref{tab:family} describe only their vote
and distance logic. Given that this memory is the largest structure in the design
space, injecting fault in it is likely the most consequential fault class we have not
measured. \emph{Hardening evaluation}, Algorithm~\ref{alg:cgsh} selects units and
predicts residual risk, and the sweep of Fig.~\ref{fig:harden} uses an area
apportionment rather than measured layout; implementing the hardened variants and
carrying them through a physical-design flow would replace the predicted budget
axis with measured silicon and allow the predicted residual risk to be verified
empirically.

\section{Conclusion}

We characterized the fault vulnerability of hardware malware detectors through $453{,}174$ structural fault evaluations, using 128 completed configurations for the primary comparisons. Tree-based detectors had a lower mean critical-fault fraction than function-based detectors, $5.7\%$ versus $14.3\%$. However, their mean metric-silent fraction was $70.2\%$, showing that stable aggregate detection metrics do not establish detector integrity. Classifier structure and fault consequences therefore provide information that fault-site count alone cannot capture.

Ensembling improved fault tolerance, but its benefits depended on the method. Bagging reduced criticality in $26$ of $27$ completed matched comparisons, whereas AdaBoost increased it in two. These results identify fault tolerance as an additional consideration when weighing the silicon cost of ensembles against their detection performance.

Selective hardening offers a complementary approach. For the representative J48 detector, protecting $16$ of $115$ structural units reduced the predicted maximum single-fault recall loss from $60.1$ to $3.3$ percentage points. This result demonstrates the potential of concentrating protection on the most consequential units under ideal correction and estimated costs. Validation against faulted RTL and measurement of physical protection overhead remain necessary to establish the practical benefits.

\end{document}